\documentstyle[11pt,fleqn]{article}   
\catcode`\@=11
\@addtoreset{equation}{section}
\def\theequation{\arabic{section}.\arabic{equation}}
\def\appendix{\renewcommand{\thesection}{\Alph{section}}\setcounter{section}{0}
              \renewcommand{\theequation}
            {\mbox{\Alph{section}.\arabic{equation}}}\setcounter{equation}{0}}
\def\maketitle{\thispagestyle{empty}\setcounter{page}0\newpage
                \renewcommand{\thefootnote}{\arabic{footnote}}
                  \setcounter{footnote}0}
\renewcommand{\thanks}[1]{\renewcommand{\thefootnote}{\fnsymbol{footnote}}
               \footnote{#1}\renewcommand{\thefootnote}{\arabic{footnote}}}
\newcommand{\preprint}[1]{\hfill{\sl preprint - #1}\par\bigskip\par\rm}
\renewcommand{\title}[1]{\begin{center}\Large\bf #1\end{center}\rm\par\bigskip}
\renewcommand{\author}[1]{\begin{center}\Large #1\end{center}}
\newcommand{\address}[1]{\begin{center}\large #1\end{center}}

\def\dinfn{\smallskip Dipartimento di Fisica, Universit\`a di Trento\\ 
                           and Istituto Nazionale di Fisica Nucleare,\\
                                   Gruppo Collegato di Trento, Italia}

\def\Idinfn{\address{\dinfn}}
\newcommand{\email}[1]{e-mail: \sl #1@science.unitn.it\rm}
\newcommand{\femail}[1]{\thanks{\email{#1}}}
\newcommand{\pacs}[1]{\smallskip\noindent{\sl PACS numbers:
                       \hspace{0.3cm}#1}\par\bigskip\rm}
\def\babs{\hrule\par\begin{description}\item{Abstract: }\it} 
\def\eabs{\par\end{description}\hrule\par\medskip\rm}
\renewcommand{\date}[1]{\par\bigskip\par\sl\hfill #1\par\medskip\par\rm}
\newcommand{\ack}[1]{\par\section*{Acknowledgments} #1} 
\newcommand{\s}[1]{\section{#1}}
\renewcommand{\ss}[1]{\subsection{#1}}

\renewcommand{\vec}[1]{{\bf #1}}       
\def\M{{\cal M}}                       
\newcommand{\ca}[1]{{\cal #1}}         
\def\hs{\qquad}               
\def\nn{\nonumber}            
\def\beq{\begin{eqnarray}}    
\def\eeq{\end{eqnarray}}      
\def\ap{\left.}               
\def\at{\left(}               
\def\aq{\left[}               
\def\ag{\left\{}              
\def\cp{\right.}              
\def\ct{\right)}              
\def\cq{\right]}              
\def\cg{\right\}}             
\def\R{{\hbox{{\rm I}\kern-.2em\hbox{\rm R}}}}   
\def\H{{\hbox{{\rm I}\kern-.2em\hbox{\rm H}}}}   
\def\N{{\hbox{{\rm I}\kern-.2em\hbox{\rm N}}}}   
\def\C{{\ \hbox{{\rm I}\kern-.6em\hbox{\bf C}}}} 
\def\Z{{\hbox{{\rm Z}\kern-.4em\hbox{\rm Z}}}}   
\def\ii{\infty}                                  
\def\X{\times\,}                                 
\newcommand{\fr}[2]{\mbox{$\frac{#1}{#2}$}}      
\def\Tr{\,\mbox{Tr}\,}                  
\def\Res{\,\mbox{Res}\,}                
\renewcommand{\Re}{\,\mbox{Re}\,}       
\def\lap{\Delta}                        
\def\al{\alpha}
\def\be{\beta}

\def\ep{\varepsilon}
\def\ze{\zeta}

\def\la{\lambda}

\def\si{\sigma}

\def\Ga{\Gamma}

\def\La{\Lambda}

\def\Om{\Omega}

\def\citen#1{%
\edef\@tempa{\@ignspaftercomma,#1, \@end, }
\edef\@tempa{\expandafter\@ignendcommas\@tempa\@end}%
\if@filesw \immediate \write \@auxout {\string \citation {\@tempa}}\fi
\@tempcntb\m@ne \let\@h@ld\relax \let\@citea\@empty
\@for \@citeb:=\@tempa\do {\@cmpresscites}%
\@h@ld}
\def\@ignspaftercomma#1, {\ifx\@end#1\@empty\else
   #1,\expandafter\@ignspaftercomma\fi}
\def\@ignendcommas,#1,\@end{#1}
\def\@cmpresscites{%
 \expandafter\let \expandafter\@B@citeB \csname b@\@citeb \endcsname
 \ifx\@B@citeB\relax 
    \@h@ld\@citea\@tempcntb\m@ne{\bf ?}%
    \@warning {Citation `\@citeb ' on page \thepage \space undefined}%
 \else
    \@tempcnta\@tempcntb \advance\@tempcnta\@ne
    \setbox\z@\hbox\bgroup 
    \ifnum\z@<0\@B@citeB \relax
       \egroup \@tempcntb\@B@citeB \relax
       \else \egroup \@tempcntb\m@ne \fi
    \ifnum\@tempcnta=\@tempcntb 
       \ifx\@h@ld\relax 
          \edef \@h@ld{\@citea\@B@citeB}%
       \else 
          \edef\@h@ld{\hbox{--}\penalty\@highpenalty \@B@citeB}%
       \fi
    \else   
       \@h@ld \@citea \@B@citeB \let\@h@ld\relax
 \fi\fi%
 \let\@citea\@citepunct
}
\def\@citepunct{,\penalty\@highpenalty\hskip.13em plus.1em minus.1em}%
\def\@citex[#1]#2{\@cite{\citen{#2}}{#1}}%
\def\@cite#1#2{\leavevmode\unskip
  \ifnum\lastpenalty=\z@ \penalty\@highpenalty \fi 
  \ [{\multiply\@highpenalty 3 #1
      \if@tempswa,\penalty\@highpenalty\ #2\fi 
    }]\spacefactor\@m}
\begin{document}

\preprint{UTF 372}
\title{
Euclidean Approach to the  Entropy \\
for a Scalar Field in Rindler-like Space-Times}

\author{
Sergio Zerbini\femail{zerbini},
Guido Cognola\femail{cognola}
and Luciano Vanzo\femail{vanzo} 
}
\Idinfn

\date{March 1996}

\babs
The off-shell  entropy for a massless scalar field in a 
D-dimensional Rindler-like space-time is investigated within the 
conical Euclidean approach in the manifold $C_\be\times\M^N$, $C_\be$ 
being the 2-dimensional cone, making use of the zeta-function regularisation. 
Due to the presence of conical singularities, it is shown that the relation 
between the zeta-function and the heat kernel is non trivial and, 
as first pointed out by Cheeger, requires a 
separation between small and large eigenvalues of the Laplace operator.
As a consequence, in the massless case, the (naive) non 
existence of the Mellin transform is by-passed by the 
Cheeger's analytical continuation of the zeta-function on manifold 
with conical singularities.
Furthermore, the continuous spectrum leads to the introduction of 
smeared traces. In general, it is pointed out that 
the presence of the divergences may depend on the smearing function 
and they arise in removing the smearing cutoff. 
With a simple choice of the smearing function, 
horizon divergences in the thermodynamical quantities are 
recovered and these are similar to the divergences found  by means of
off-shell methods like the brick wall model, the optical conformal 
transformation techniques or the canonical path integral method.
\eabs


\s{Introduction}
As is well known there exist several methods for 
calculating the semiclassical entropy (tree-level contribution) 
for a stationary black hole (see for example \cite{wald95u-24}). 
In the Einstein theory, for a non rotating 4-dimensional black hole, 
all the methods lead to the celebrated  Bekenstein-Hawking classical 
entropy $4\pi GM^2$  
\cite{beke73-7-2333,hawk75-43-199,gibb77-15-2752}. 
The thermodynamical origin of 
this quantity is well known and recently in a series of  
papers (see for example \cite{frol95u-184} and references therein) 
this fact has been stressed, namely the Bekenstein-Hawking entropy can be 
defined by the response of the free energy of the black hole to the 
change of the equilibrium (Unruh-Hawking) temperature. This temperature 
depends on the mass of the black hole and may be determined by requiring the 
smoothness of the related Euclidean solution 
\cite{gibb77-15-2752}. 
This is an example of an on-shell computation. In the derivation 
of the above result, one usually neglects quantum fluctuation effects. 
If one takes quantum effects into account, one can show 
that the on-shell one-loop contribution is finite 
(see for example  Ref. \cite{frol95u-184}).

The situation drastically changes if one tries to investigate the  
issue of the black hole entropy within a statistical-mechanical 
approach, i.e. by counting the quantum states of the black hole. 
In this case, in order to evaluate the entropy, one is 
forced to work off-shell, namely at a temperature different from the 
Unruh-Hawking one. The  first off-shell computation of 
the black hole entropy has been appeared in the  't~Hooft seminal paper 
\cite{thoo85-256-727}, where the black hole degrees of freedom have 
been identified with the ones of a quantum gas of scalar particles 
propagating outside, but very near the horizon at a temperature $\be^{-1}$. 
The statistical-mechanical quantities were found to be divergent 
and regularised by Dirichlet boundary conditions imposed at a small 
distance from the black hole horizon (the so called brick-wall model).
  
In a generic off-shell procedure, these divergences are not totally unexpected. 
In fact their physical origin may be described by the following simple 
considerations. The equivalence principle implies that a system in thermal 
equilibrium has a local Tolman temperature  given by 
$T(x)=T/{\sqrt {|g_{00}(x)|}}$, $T$ being the temperature measured at 
the spatial infinity. Thus, the asymptotic high 
temperature expansion for the free energy of a massless quantum gas
on a $D$-dimensional static space-time may be written as 
\beq
F_T\simeq
T^{D}\int |g_{00}(\vec{x})|^{-D/2} 
\sqrt{|g_{D-1}(\vec{x})|}\,dx^{D-1}  
\:,\nn\eeq
where $g_{D-1}=\det\{g_{ij}\}$ ($i,j=1,...,D-1$).
In the presence of horizons, the integrand have non summable singularities 
and horizon divergences appear. As a consequence also the entropy is 
divergent. The nature of these divergences depends on the 
zeros and the poles of $g_{00}$ and $g_{D-1}$ respectively. 
In general, for extreme black holes, where  $g_{00}$ has higher order 
zeros, the divergences are much more severe than the divergences 
in the non extremal case (see for example \cite{cogn95-52-4548}). 
 
These considerations suggest the use of another off-shell method, 
based on conformal transformation techniques, which consists in 
mapping the original metric onto the optical one 
$\bar{g}_{\mu\nu}=g_{\mu\nu}/g_{00}$, 
(see Refs.~\cite{dowk78-11-895,page82-25-1499,brow85-31-2514}).  
Related methods which lead to optical manifolds have  been considered 
in Refs. \cite{barb94-50-2712,barb95-52-4527,deal95-52-3529}.
The conformal optical transformation method has been used
in the case of fields in 4-dimensional black hole space-time 
\cite{barv95-51-1741,cogn95-12-1927} and also for massive scalar fields in 
$D$-dimensional Rindler-like space-times \cite{byts96-458-267}.  
These are  space-times of the form  
$\R\times\R^+\times\M^N$, with metric 
\beq
ds^2=-\frac{b^2\rho^2}{r_H^2} dx_0^2+d\rho^2+d\si^2_N
\:,\hs N=D-2\:,
\label{rl}
\eeq
where $r_H$ is a dimensional constant, $b$ a constant factor and 
$d\si^2_N$ the spatial metric related to the smooth manifold $\M^N$. 
If $\M^N=\R^N$, then $b=1$, $r_H=1/a$, $a$ being the constant acceleration, 
the manifold is non compact and one has to deal 
just with the $D$-dimensional Rindler space-time. 
If  $\M^N=S^N$, one can show that one is dealing with a space-time 
which approximates, near the horizon and in the large mass limit, 
a $D$-dimensional Black hole (see, for example \cite{byts96-458-267}). 
In this case $b=(D-3)/2$ and $r_H$ is the horizon radius
depending on the mass of the black hole. 

In the case of canonical horizons 
(this means that $g_{00}(\vec{x})$ has simple zeros only) 
and in order to treat finite temperature effects,
an alternative off-shell approach has been proposed 
\cite{suss94-50-2700,call94-333-55,solo95-51-618,furs95-10-649,kaba95-453-281}. 
It consists in performing the Wick rotation $x_0=-i\tau$ 
and working in the Euclidean continuation of the space-time, 
with the imaginary time compactified to an arbitrary interval of 
length $\be$ and with the fields 
periodic in $\tau$ with period $\be$, which is interpreted as the 
inverse temperature. 
For an arbitrary choice of $\be$, such a static manifold
has a conical-like singularity. As already mentioned, 
only in the absence of such a singularity there exist equilibrium 
states with finite stress-energy tensor. This happens only for 
$\be=\be_H$, the Unruh-Hawking temperature \cite{gibb77-15-2752}.
Furthermore, the Bekenstein-Hawking entropy can be derived within this 
approach \cite{bana94-72-957,suss94-50-2700}. 
For the metric of our simplified model, Eq.~(\ref{rl}), one has
\beq 
\be_H=\frac{2\pi r_H}{b}\,.
\label{ht}
\eeq

The situation looks quite different in the computation of
the quantum corrections to the entropy. In fact in this case  
one is really forced to consider an arbitrary $\be$, thus working off-shell 
again. This proposal seems highly non trivial, 
since only in an ultrastatic space-time the imaginary 
time formalism has been shown to be equivalent to the canonical 
formalism of the finite temperature quantum field theory. 
The use of the trace of the heat-kernel plus 
standard proper-time regularisation 
on manifolds with conical singularities leads to a free energy, 
which is ultraviolet divergent
and with a dependence on temperature different from the one expected for a 
$D$-dimensional space. In fact the leading term goes as $T^2$
independently on the dimensions \cite{empa95-51-5716}.
Let us summarise this crucial issue, which seems to have been 
overlooked in the recent papers.  

It is well know (see Sec.~2) that one-loop effects can be described by 
zeta-function regularisation. The zeta-function related to a free 
massless scalar field on $ \M^D$ can be obtained by means of the  
Mellin transform of $K(t|-\lap_D)=\Tr\exp(t\lap_D)$, $\lap_D$ being  the Laplace 
operator.  However,
in the non compact manifold $\M^D=C_\be\times\M^N$, 
in order to give meaning to the trace of the heat-kernel, one has to use a smearing function $\phi$. 
A simple choice is given by  the product of the characteristic function 
$\chi(\Om)$ ($\Om\subset\R^N$, compact) 
and a cutoff function $\theta(\La-\rho)$ regularizing the infinite conical volume. 
One has
\cite{chee83-18-575,dowk87-36-620,cogn94-49-1029,furs94-11-1431}
\beq
K(t|-\lap_D)(\phi)=\frac{V_D}{(4\pi t)^{\frac D2}}
+\frac1{12}\at\frac{2\pi}{\be}-\frac{\be}{2\pi}\ct
\:\:\frac{V_N}{(4\pi t)^{\frac N2}}
\:,\nn\eeq
$V_N$ being the volume of $\Om$ and $V_D=\be\La^2V_N/2$. As a result, 
the naive Mellin transform of this heat-kernel trace does not exist or is zero if 
interpreted in the sense of distributions.  
If a mass term is included, then the latter equation has to be 
multiplied by the factor $\exp(-tm^2)$ and the global zeta-function 
may be defined via the Mellin transform \cite{cogn94-49-1029}, 
obtaining in this way a well defined quantity, 
apart the volume divergences associated with the non compactness of the 
manifold.
The surprising thing is that, in contrast with the other methods
mentioned above, such a partition function seems not to have any horizon 
divergence. It has also a dependence on $\be$, which does not depend on 
the dimensions of the Rindler space one is considering and, besides,   
it vanishes in the limit $m\to0$. 
Of course, the contribution computed in this way is only the  finite part of 
the partition function, 
since ordinary ultraviolet divergences are present (formally one is 
dealing with a "zero-temperature" field theory on a non trivial 
background) and they have been 
accounted for by means of the zeta-function regularisation. 
If one makes use of another regularisation, 
for example the proper-time regularisation, it 
turns out that such ultraviolet divergences are not confined to the 
vacuum sector as in the usual finite-temperature theory on ultrastatic 
space-times, but they appear also
in the non trivial part of the free energy  
in such a way that they give a contribution to the entropy, 
even if it is evaluated at the Hawking temperature 
\cite{call94-333-55,solo95-51-618,furs95-10-649,kaba95-453-281}. 
In the limit $m\to0$ only the leading divergent term remains and this 
has been  interpreted as the analogous of the horizon divergence.  

In this paper, in order to try to elucidate this issue, we will present  an off-shell Euclidean 
zeta-function regularisation approach applicable directly in the massless case 
and we will make the comparison between this Euclidean 
conical method and the optical conformal transformation methods 
(the brick wall method gives the same result as the latter). 
To avoid the risk to create confusion,
other kind of regularisations shall not be taken into account.
Our main aim will be the construction of the zeta-function for a massless 
scalar field in $C_\be\times\M^N$, starting from the local one, which 
can be evaluated  
by using  an analytical procedure suggested by Cheeger \cite{chee83-18-575}. 
As  mentioned above, all global quantities, for example free energy 
or entropy, which are related to the trace of some operator, 
require a smearing function in order to be defined. 
The horizon divergences appear in the smearing removal. 
It turns out that the change of the smearing prescription 
may modify the final resul. 
However we will show that the results obtained in this way are compatible with 
the ones obtained by using the optical conformal  method, including 
the correct dimensional behaviour of the free energy on $\be$.
We also would like to mention that the techniques presented in this 
paper may be useful in investigating quantum fields in space-times 
with spatial conical singularity (see, for example, 
\cite{furs94-11-1431,more95-451-365}  and references therein).

The contents of the paper are the following. 
In Sec.~2, the general formalism is summarised and the partition function, 
as well as the other related quantities, are introduced. 
In Sec.~3 the local zeta-function for a massless scalar field in a Rindler 
space is constructed according the Cheeger's method and 
the global (smeared) zeta-function is computed and then generalised to any
Rindler-like space-times.
In Sec.~4 the thermodynamical quantities in such kind of spaces
are derived and their properties analysed. 
The application to the 4-dimensional black hole, in the large mass 
limit approximation, is presented in Sec.~5.
Choosing a suitable smearing function, horizon divergences of the 
first quantum corrections to the free energy are recovered. 
The conclusions are reported in Sec.~6.
The paper ends with an Appendix devoted to the analytical extension of a 
series which frequently appears in the formulae and which plays an 
important role in the Cheeger method.

\s{General formalism}
\label{Form}

To start with we recall the formalism we shall use in the following in 
order to discuss the finite temperature effects within this conical 
singularity approach. We may consider, as a 
prototype of the quantum correction 
(quantum degrees of freedom), a scalar field on a $D$-dimensional  
Rindler-like space-time. Its related Euclidean metric reads
\beq
ds^2=\frac{b^2\rho^2}{r_H^2} d\tau^2+d\rho^2+d\si^2_N\,,
\hs x=(\tau,\rho,\vec{x})\,,
\nn\eeq
where $\tau$ is the imaginary time, $\rho \geq 0$ the radial coordinate 
and $\vec x$ the transverse coordinates. 
As explained in the introduction, finite 
temperature effects are assumed to arise when $\tau$ is compactified  
$0\leq\tau\leq\be$, $\be$ being the inverse of the temperature. 
For arbitrary $\be$ the manifold $\M^D$ has the topology of 
$C_\be\times\M_N$, $C_{\be}$ being the 2-dimensional cone, 
$(\tau,\rho)\in C_{\be}$,  $\vec x\in\M_N$. 
From now on, we put $r_H=1$ and $\tau\to b\tau$.
 
The one-loop partition function depends on $\be$ and is given by 
\begin{equation}
Z_\be=\int d[\phi]\,
\exp\at-\frac12\int\phi L_D \phi d^Dx\ct
\:,\end{equation}
where $\phi$ is a scalar density of weight $-1/2$, which obeys 
periodic boundary conditions $\phi(0,\vec x)=\phi(\be,\vec x)$  
and $L_D$ is the Laplace-like operator on $C_{\be}\times\M_N$.
In our case, it has the form
\beq
L_D=-\lap_D+\xi R+m^2=-\lap_\be+L_N=-\lap_\be-\lap_N+\xi R+m^2
\:.\eeq
Here $\lap_D$, $\lap_N$ and $\lap_\be$ are the Laplace-Beltrami 
operators on $\M^D$, $\M^N$ and $C_\be$ respectively,
$\xi$ is an arbitrary parameter,  $m$ the mass
and $R$ the scalar curvature of the manifold, 
which is assumed to be a constant. 

In the one-loop or external field approximation
the importance of the zeta-function regularisation as a
powerful tool to deal with the ambiguities (ultraviolet divergences) present 
in the relativistic quantum field theory 
is well known (see for example \cite{eliz94b}). 
It permits to give a meaning, in the sense of analytic continuation, 
to the determinant of a differential operator which, as product of 
eigenvalues, is formally divergent.    
One has \cite{hawk77-55-133}
\beq
\ln Z_\be=-\frac{1}{2}\ln\det L_D
=\frac{1}{2}\ze_\be'(0|L_D)+\frac{1}{2}\ze_\be(0|L_D)\ln\mu^2
\nn\:,\eeq 
where $\ze_\be(s|L_D)$ is the zeta-function related to $L_D$,
$\ze_\be'(0|L_D)$ its derivative with respect to $s$ and 
$\mu^2$ a renormalisation scale.
The analytically continued zeta-function is regular at $s=0$ 
and thus its derivative is well defined.

When the manifold is smooth and compact the spectrum is discrete and one has
\beq
\ze_\be(s|L_D)=\sum_i\la_i^{-2s}
\:,\nn\eeq
$\la^2_i$ being the eigenvalues of $L_D$. As a result,
one can make use of the relationship between the zeta-function and the 
heat-kernel trace via the the Mellin transform and its inverse. 
For $\Re s>D/2$, one can write
\beq
\ze_\be(s|L_D)=\Tr L_D^{-s}
=\frac{1}{\Ga(s)}\int_0 ^\ii t^{s-1} 
\:K_\be(t|L_D)\:dt
\,,\label{mt}\eeq
\beq
K_\be(t|L_D)=\frac{1}{2\pi i}
\int_{\Re s >D/2} t^{-s}\:\Ga(s)\ze_\be(s|L_D)\:ds
\:,\label{minv}
\eeq
where $K_\be(t|L_D)=\Tr\exp(-t L_D)$ is the heat operator.
The previous relations are valid also in the presence of 
zero modes with the trivial replacement 
$K_\be(t|L_D)\longrightarrow K_\be(t|L_D)-P_0$,
$P_0$ being the projector onto the zero modes. 
We may call this  the global approach. 
Moreover one may follow a local 
approach, starting from local quantities like the heat-kernel and the 
related Mellin transform local zeta-function $\ze_\be(s;x|L_D)$. 
Then one may introduce an effective Lagrangian density
\beq
\ca{L}(x)=\frac{1}{2}\ze_\be'(0;x|L_D)
+\frac{1}{2}\ze_\be(0;x|L_D)\ln\mu^2
\:,\nn\eeq
obtaining in this way
\beq
\ln Z_\be=\frac{1}{2}\int 
\aq \ze_\be'(0;x|L_D)+\ze_\be(0;x|L_D,0)\ln\mu^2\cq\:dV_D 
\,.\label{zeta1}\eeq 
Normally the two approaches give the same results. In the presence of 
conical singularities and in the massless case, we have seen that the 
global approach cannot be used, so we shall make use of the local one.
In the presence of conical singularities and in the non compact case 
(continuos spectrum),
some care has to be used in the implementation of the 
relationship between heat-kernel and local zeta-function. 
With regard to this, we shall show in the next Section that a 
separation of lower eigenvalues from the higher ones, 
together with  a suitable analytic continuation is necessary 
\cite{chee83-18-575}. 
Furthermore the local zeta-function turns out to be a non local 
summable function. For this reason one can take the distributional 
characters of the local zeta-function into account by introducing a 
smearing in terms of a function with compact support. In this way, in 
order to treat global quantities, one has to deal with smeared traces 
\cite{brun85-58-133}. 

Once the (smeared) partition function is given by 
Eq.~(\ref{zeta1}), we assume the validity of the usual thermodynamical 
relations, thus the free energy can be computed by means of 
\beq 
F_\be=-\frac{1}{\be}\ln Z_\be
=-\frac{\ze_\be'(0|L_D)}{2\be}
-\frac{\ze_\be(0|L_D)}{2\be}\:\:\ln\mu^2 
\:,\label{FE}\eeq
and the entropy and the internal energy read
\beq
S_\be=\be^2 \partial_\be F_\be \:,
\hs U_\be=\frac{S_\be}{\be}+ F_\be\:.
\label{entropy}\eeq
All these quantities are evaluated off-shell. However the only admissible 
equilibrium thermal state is the one corresponding to the Unruh-Hawking 
temperature $\be=\be_H$ (in our model, see Eq.~(\ref{ht})). Thus, strictly speaking, 
one has to make the comparison of the different off-shell approaches only at $\be=\be_H$.

\s{Zeta-function regularisation in a space with conical singularities}

As we have mentioned in the introduction, the evaluation of the partition 
function using the zeta-function regularisation requires some care. 
Here we shall evaluate the kernel of the zeta-function and then 
we will give a meaning to the global zeta-function by means of smearing. 
As has been stressed by Cheeger \cite{chee83-18-575}, it is crucial to treat small and large 
eigenvalues separately.

The spectral properties of the Laplace operator on the cone are well known
and in fact, a complete set of normalised eigenfunctions 
for $L_\be=-\lap_\be$ is easily found to be  
\begin{equation}
\psi(\tau,\rho)=\frac{1}{\sqrt{2\pi}}
e^{i\nu_l\tau}J_{\nu_l}(\la\rho)
\:,\hs\nu_l=\frac{2\pi l}\be\:,\hs l\in\Z
\:,\end{equation}
together with its complex conjugate (double degeneration).
Here $\la^2$ ($\la\geq0$) is the eigenvalue corresponding to $\psi$
and $\psi^*$, while $J_\nu$ is the regular Bessel function.
This choice of the eigenfunctions correspond to a
positive elliptic self-adjoint operator 
(the Friedrichs extension \cite{brun87-73-369,kay91-139-103}).

Now, using the standard separation of variables, it is easy 
 to get the spectrum and the eigenfunctions of the operator 
$L_D=-\lap_\be+L_N$ on the Rindler-like space-time 
$\M^D=C_\be\X\M^N$, $L_N$ being a Laplace type operator on 
$\M_N$ including (eventually) mass and scalar curvature coupling term. 
Indicating by $f_\al(\vec x)$ and $\la^2_\al$ respectively the 
eigenvectors and the eigenfunctions of $L_N$, one has 
$\Psi(x)=\psi(\tau,\rho)f_\al(\vec x)$ and $\la^2+\la_\al^2$
for the eigenvectors and the eigenfunctions of $L_D$. 
Thus, for the diagonal kernel of a operator  $F_\be(L_D)$ one has
\beq
F_\be(x|L_D)=
\frac1{\be}\sum_\al\aq
\int_{0}^{\infty}F(\la^2+\la^2_\al)
J^2_0(\la\rho)\la\:d\la+
2\sum_{l=1}^{\ii}
\int_{0}^{\infty}F(\la^2+\la^2_\al)
J^2_{\nu_l}(\la\rho)\la\:d\la
\cq\:.
\label{spec}\eeq
As it stands, such an expression is only formal, since 
the series and the integral could not be convergent.

\ss{A special case: massless scalar field in Rindler space}

For the sake of simplicity and for illustrative purposes, 
let us start to consider a massless scalar field on Euclidean Rindler 
space $C_\be\times\R^N$. We suppose $N\geq1$, but all results on 
the pure cone ($N=0$) can be obtained as limit cases.
For this case  $L_N=-\lap_N$ has a continuous spectrum 
$\la^2_{\vec k}=k^2$ and so the sum over $\al$ reduces to an integral 
over $\vec k\in\R^N$ and its spectral data are well known, namely
\beq
f_{\vec k}=\frac{e^{i\vec k\cdot\vec x}}{(2\pi)^{N/2}}
\:,\hs \la^2_{\vec k}=k^2=\vec k \cdot \vec k
\:.\nn\eeq

We are interested in the local zeta-function, so we choose
$F(L_D)=L_D^{-s}$ and, using Eq.~(\ref{spec}), one formally has
\beq
\ze_\be(s;x|L_D)&=&
\frac{2(4\pi)^{-\frac N2}}{\be\Ga(\frac N2)}
\int_{0}^{\ii}dk\: k^{N-1}\aq
 \int_{0}^{\infty}(\la^2+k^2)^{-s}
J^2_0(\la\rho)\la\:d\la
\cp\nn\\&&\ap\hs\hs\hs
+2\sum_{l=1}^{\ii}
\int_{0}^{\infty}(\la^2+k^2)^{-s}
J^2_{\nu_l}(\la\rho)\la\:d\la
\cq\:.\label{vb}\eeq
Recalling the asymptotic behaviour of Bessel functions one can easily
see that both the integrations over $k$ and $\la$ can be performed
in any term of the latter equation if $s$ is restricted in the range 
\beq
\frac{N+1}2<\Re s<\frac{N+2+2\nu_l}2
\:.\nn\eeq
In fact one has
\beq
\frac{2(4\pi)^{-\frac N2}}{\be\Ga(\frac N2)}
\int_{0}^{\ii}dk\: k^{N-1}
 \int_{0}^{\infty}(\la^2+k^2)^{-s}
J^2_{\nu_l}(\la\rho)\la\:d\la
&=&\nn\\&&\hspace{-5cm}
\frac{\rho^{2s-D}}{2\be(4\pi)^{\frac N2}\Ga(s)}
\:\frac{\Ga(s-\fr{N+1}2)\Ga\at\nu_l-(s-\fr N2)+1\ct}
{\sqrt{\pi}\Ga(\nu_l+s-\fr N2)}
\:.\label{Nuova}\eeq
To get the zeta-function, now one has to sum over $l$. As we shall 
show in the Appendix, the series is convergent for $\Re s>\frac N2+1$. 
This range does not overlap with the previous one for 
$\nu_l=0$ ($l=0$). This means that there are no values of $s$ for which
Eq.~(\ref{vb}) is a finite quantity.
The solution of this convergence obstruction has been 
suggested by Cheeger \cite{chee83-18-575}. It simply consists in a 
separate treatment of the lower and the higher eigenvalues 
(in this particular case $\nu_0=0$ and $\nu_l>0$, $l\geq1$).
Only after the analytic continuation is performed, one may define the zeta-function by 
summing the two contributions obtained in this way. Of 
course, such a definition of zeta-function has all the requested 
properties and coincides with the usual one when the manifold is smooth.

So, following Cheeger, in Eq.~(\ref{vb}) we first isolate the term $l=0$ 
and define (see Eq.~(\ref{Nuova})) for $\frac12+\frac{N}2<\Re s<1+\frac{N}2$ 
\beq
\ze_<(s;x|L_D)&=&
\frac{\rho^{2s-D}}{\be(4\pi)^{\frac{N}{2}}\Ga(s)}
\:\frac{\Ga(s-\fr{N+1}2)\Ga(1-s+\fr N2)}
{2\sqrt{\pi}\Ga(s-\fr{N}{2})} \,,\nn\\
&=&-\frac{\rho^{2s-D}}{\be(4\pi)^{\frac{N}{2}}\Ga(s)}
\:\frac{\Ga(s-\fr{N+1}2)G_{2\pi}(s-\fr N2)}
{\sqrt\pi} \,.\label{mnlow}
\eeq
Then we consider all the other terms, perform the integration 
as in Eq.~(\ref{Nuova}) and the summation over $l\geq1$. 
In this second case we have to restrict to 
$1+\frac N2<\Re s<1+\nu_1+\frac{N}2$. The result reads 
\beq
\ze_>(s;x|L_D)t
=\frac{\rho^{2s-D}}{\be(4\pi)^{\frac N2}\Ga(s)}
\,\frac{\Ga(s-\fr{N+1}2)G_\be(s-\frac N2)}
{\sqrt\pi} \,.
\label{mnmr2}
\eeq
We have put
\beq
G_\be(s)=\sum_{l=1}^\ii 
\frac{\Ga(\nu_l-s+1)}{\Ga(\nu_l+s)} \,, 
\hs G_{2\pi}=-\frac{\Ga(1-s)}{2\Ga(s)}\:,
\nn\eeq
the series being convergent for $\Re s>1$. As we shall show in the 
Appendix, the analytic continuation of $G_\be(s)$ is a meromorphic 
function with only a simple pole at $s=1$. 
This means that both Eqs.~(\ref{mnlow}) and (\ref{mnmr2}) can be 
analytically continued to the whole complex $s$ plane and, 
by definition 
\beq
\ze_\be(s;x|L_D)=
\ze_<(s;x|L_D)+\ze_>(s;x|L_D)
=\frac{\rho^{2s-D}}{\be(4\pi)^{\frac N2}\Ga(s)} 
\:I_\be(s-\fr{N}2)\:,
\label{mn1}\eeq
where
\beq
I_\be(s)=\frac{\Ga(s-\fr12)}{\sqrt{\pi}}
\aq G_\be(s)-G_{2\pi}(s)\cq
\:.\nn\eeq
The properties of $G_\be$, as well as of $I_\be$, will be studied in 
the Appendix. An important property is that $I_\be$, as well as
$G_\be$, has only a simple pole at $s=1$. 

Note that in spite of the definition (\ref{mn1}),
in the use of the inverse Mellin transform 
one has to consider $\ze_<$ and $\ze_>$
again separately and the original ranges of convergence.
That is
\beq
K_<(t;x|L_D)
&=&\frac1{2\pi i}\int_{\frac12+\frac{N}2<\Re s<1+\frac{N}2} 
\:\:\: t^{-s}\Ga(s) \ze_<(s;x|L_D)\:ds\:,\label{M0}\\
K_>(t;x|L_D)
&=&\frac1{2\pi i}\int_{1+\frac N2<\Re s<1+\nu_1+\frac{N}2} 
\:\:\: t^{-s}\Ga(s) \ze_>(s;x|L_D)\:ds
\:,\label{M1}\\
\ze_<(s;x|L_D)&=&\frac1{\Ga(s)}
\int_{0}^{\ii}t^{s-1}K_<(t;x|L_D)\:dt\:,
\hs\fr12+\fr{N}2<\Re s<1+\fr{N}2\:,\label{AM0}\\
\ze_>(s;x|L_D)&=&\frac1{\Ga(s)}
\int_{0}^{\ii}t^{s-1}K_>(t;x|L_D)\:dt\:,
\hs1+\fr{N}2<\Re s<1+\nu_1+\fr{N}2
\:,\label{AM1}\eeq
and, by definition
\beq  
K_\be(t;x|L_D)&\equiv &K_<(t;x|L_D)+K_>(t;x|L_D)
\nn\\&=&
\frac{2\pi}{\be}\frac1{(4\pi t)^{\frac D2}}
+\frac{(4\pi)^{-\frac N2}}{2\pi i\be}
\int_{\Re s>1+\frac{N}2} 
\:\:\: t^{-s}\rho^{2s-D}I_\be(s-\fr N2)\:ds
\:.\label{Mell}
\eeq
On the right hand side of the latter equation we immediately recognise
the kernel $K_{2\pi}(t;x|L_D)$ and so the integral represents 
the difference
$K_\be(t;x|L_D)-\frac{2\pi}{\be}K_{2\pi}(t;x|L_D)$.
Similar expressions are valid for all quantities. 
This can be seen by observing that 
Eq.~(\ref{spec}) can be written in the form 
\beq
F_\be(x|L_D)-\frac{2\pi}{\be}F_{2\pi}(x|L_D)=
\frac2\be\sum_\al\ag \sum_{l=1}^{\ii}
\int_{0}^{\infty}F(\la^2+\la^2\al)
\aq J^2_{\nu_l}(\la\rho)-J^2_l(\la\rho)
\cq \la\:d\la \cg\:.
\nn\eeq
The advantage is that the low eigenvalue $\nu_0=0$ 
is absent on the right-hand side of this expression.  
As a result
\beq
\ze_\be(s;x|L_D)-\frac{2\pi}{\be}\ze_{2\pi}(s;x|L_D)&=&
\frac{4(4\pi)^{-\frac N2}}{\Ga(\frac N2)}
\int_{-\ii}^{\ii}dk\: k^{N-1}\nn\\ &\times& 
\sum_{l=1}^{\ii}\int_{0}^{\infty}(\la^2+k^2)^{-s}
\aq J^2_{\nu_l}(\la\rho)-J^2_l(\la\rho)\cq
\la\:d\la
\:.\nn\eeq
Now the right hand side of the latter equation is well defined for
$1+\frac N2<\Re s<1+\nu_1+\frac{N}2$. After integration one has 
\beq
\ze_\be(s;x|L_D)
-\frac{2\pi}{\be}\ze_{2\pi}(s;x|L_D)
=\frac{\rho^{2s-D}}{\be(4\pi)^{\frac N2}\Ga(s)} 
\:I_\be(s-\fr{N}2)\,,
\nn\eeq
which is identical to the previous definition of zeta-function
and this means that, for this particular case, the Cheeger analytical 
procedure gives $\ze_{2\pi}(s;x|L_D)=0$ 
(note that formally $\ze_{2\pi}$ is a divergent integral 
whatever is $s$).

Heat kernel and local zeta-function are related by
\beq
\ze_\be(s;x|L_D)&-&\frac{2\pi}{\be}\ze_{2\pi}(s;x|L_D)
\nn\\&=&
\frac1{\Ga(s)}\int_0^\ii\:t^{s-1}\aq
K_{\be}(t;x|L_D)-\frac{2\pi}{\be}K_{2\pi}(t;x|L_D)
\cq\:dt
\:,\label{Mellin1}\eeq
\beq
K_{\be}(t;x|L_D)&-&\frac{2\pi}{\be}K_{2\pi}(t;x|L_D)
\nn\\&=&
\frac1{2\pi i}\int_{\Re s>1+\frac{N}2} 
t^{-s}\Ga(s)\aq \ze_\be(s;x|L_D)
-\frac{2\pi}{\be}\ze_{2\pi}(s;x|L_D)\cq\:ds
\:.\label{Mellin2}\eeq
Note that for $\be=2\pi$ the conical singularity disappears and the 
manifold becomes $\R^D$. Thus, $\ze_{2\pi}$
and $ K_{2\pi}$ are trivial. Furthermore, by making use of 
Eq.~(\ref{Mell}) and taking the analytical properties of $I_\be(s)$ 
discussed in the Appendix into account, one gets the asymptotics of 
the heat-kernel, namely
\beq  
K_\be(t;x|L_D) \simeq \frac1{(4\pi t)^{\frac D2}}
+E_t(\rho)
\:.\label{Mell2}
\eeq
where $E_t(\rho)$ is an exponentially small term in $t$. This local 
asymptotics is in agreement with the results of Refs.
\cite{chee83-18-575,dowk87-36-620,cogn94-49-1029,furs94-11-1431}.

We conclude this Section introducing the global quantities.
Strictly speaking, only the distributional trace has a mathematical 
meaning, since the local zeta-function above has non integrable 
singularities in $\rho$ (see for example \cite{brun85-58-133}). 
As a consequence one has to introduce a smearing 
by means of a suitable function $\phi(\rho)$ with 
compact support not containing the origin, thus defining
\beq
\ze_\be(s|L_D)(\phi)=\be \int dV_N\int_0^\ii \phi(\rho)
\ze_\be(s;x|L_D)\rho \, d\rho
\,.\label{rgtr}
\eeq
For the smeared trace we get
\beq
\ze_\be(s|L_D)(\phi)
=\frac{V_N}{(4\pi)^{\frac{N}{2}}}
\:\frac{I_\be(s-\frac{N}{2})\hat{\phi}(2s-N)}{\Ga(s)}
\,,\hs
\hat{\phi}(s)=\int_0^\ii \rho^{s-1} \phi(\rho) d\rho
\:,\eeq
$\hat\phi$ being an analytic function since the integral in
Eq.~(\ref{rgtr}) exists for all $s$ by definition.
As a smearing function we may  simply choose $\phi(\rho)=\theta(\La-\rho)\theta(\rho-\ep)$
($\La>\ep$), which is convergent to 1 in the limits
$\La\to\ii$ and $\ep\to0$. Thus, we have
\beq
\hat\phi(s)=\frac{\La^s-\ep^s}{s}
\nn\eeq
and for the smeared zeta-function 
\beq
\ze_\be(s|L_D)(\phi)
=\frac{V_N}{(4\pi)^{\frac N2}}
\:\frac{I_\be(s-\frac N2)(\La^{2s-N}-\ep^{2s-N})}
{\Ga(s)(2s-N)}\:,\hs
\ze_\be(0|L_D)(\phi)=0
\,.\label{mnmr7}\eeq

\ss{The general case: scalar fields in Rindler-like spaces}

Here we will give some results concerning the more general case
$\M^D=C_\be\times\M^N$, $\M^N$ being an arbitrary smooth manifold
without boundary. 
In such conditions all known results concerning heat-kernel and 
zeta-function for $L_N$ on $\M^N$, which we suppose to be known,
are applicable. In particular we remind that the kernels 
are related by means of Mellin transforms, the analogues of 
Eqs.~(\ref{mt}) and (\ref{minv}). 
Furthermore, the heat kernel has the usual asymptotic expansion
\beq
K(t;\vec x|L_N)\simeq\sum_r A_r(\vec x|L_N) 
t^{r-\frac{N}2}\,,
\nn\eeq
while the local zeta-function has the meromorphic structure 
(theorem of Seeley)
\beq
\Ga(s)\ze(s;\vec x|L_N)= 
\sum_r \frac{A_r(\vec x|L_N)}
{s+r-\frac{N}2}+\mbox{ the analytical part}\,,
\label{seel}\eeq
the spectral coefficients $A_r(\vec x|L_N)$ being computable functions
(for a review see \cite{bran90-15-245}). 
Here we suppose zero-modes to be absent, but of course one can take 
them into account with simple modifications of the formulae.
 
Now let us try  to derive the meromorphic structure 
of $\ze_\be(s;\vec x|L_D)$ on 
$\M^D$. To this aim we use the factorisation 
property of the heat-kernel 
\beq
K_\be(t;x|L_D)=K(t;\tau,\rho|L_\be)K(t;\vec x|L_N)
\:,\label{fact}\eeq
in which the heat kernels of the 
Laplace-like operators on $\M^D$, $C_\be$ and $\M^N$ respectively appear.
By taking the Mellin transform of Eq.~(\ref{fact}) one usually gets 
the Dikii-Gelfand representation for the zeta-function, 
which easily permits to read off the meromorphic structure. 
However, as we have shown in the previous section, in the presence of 
conical singularities we have to separate low and high eigenvalues
in order to have a well defined Mellin transform. So we set
\beq
\ze_<(s;x|L_D)&=&\frac{1}{\Ga(s)}
\int_0^\ii t^{s-1}K_<(t;\tau,\rho|L_\be)K(t;\vec x|L_N)\,dt
\:,\nn\\
\ze_>(s;x|L_D)&=&\frac{1}{\Ga(s)}
\int_0^\ii t^{s-1}K_>(t;\tau,\rho|L_\be)K(t;\vec x|L_N)\,dt
\:,\nn\eeq
where $K_<(t;\tau,\rho|L_\be)$ and $K_>(t;\tau,\rho|L_\be)$
are related to the corresponding zeta-functions by mean of 
Eqs.~(\ref{M0}-\ref{AM1}), but with $N=0$ (pure cone). 
Now we make use of  the  Mellin-Parseval identity
\beq
\int_{0}^{\ii}f(t)g(t)dt
=\frac1{2\pi i}\int_{\Re z=c}\hat f(z)\hat g(1-z)\:dz
\:,\eeq
where $c$ is in the common strip of analyticity of the Mellin transforms 
$\hat f(z)$ and $\hat g(1-z)$.
After some calculations paying attention to the range of convergence we get
\beq
\ze_<(s;x|L_D)&=&
\frac1{2\pi i\Ga(s)}\int_{c_0}
\Ga(z)\ze_<(z;\tau,\rho|L_\be)
\Ga(s-z)\ze(s-z;x|L_N)\:dz
\:,\label{ze--}\\
\ze_>(s;x|L_D)&=&
\frac1{2\pi i\Ga(s)}\int_{c_1}
\Ga(z)\ze_>(z;\tau,\rho|L_\be)
\Ga(s-z)\ze(s-z;x|L_N)\:dz
\:,\label{ze++}\eeq
where $\frac12<c_0<1$ and $1<c_1<\Re s-\frac N2$.
These are the Dikii-Gelfand representations for  $\ze_<$ and
$\ze_>$ which are valid for $\Re s>1+\frac N2$. Since they are well
defined in the same range we can directly write the 
zeta-function as $\ze=\ze_<+\ze_>$. Then, a representation
valid for $\Re s>1+\frac N2$ reads
\beq
\ze_\be(s;x|L_D)&=&\frac{\ze(s-1;\vec x|L_N)}{2\be(s-1)}
\nn\\&&+\frac1{2\pi i\be\Ga(s)}\int_{\Re z=c_1}
\rho^{2z-2}I_\be(z)
\Ga(s-z)\ze(s-z;x|L_N)\:dz
\:,\label{zzz}\eeq
where in Eq.~(\ref{ze--}) we have shifted
the contour integral on the right,  taking into account that the integrand 
function has a simple pole at $z=1$ (see Eq.~(\ref{mnlow})). 
We incidentally observe that the first term on the right-hand side
of the latter equation is just $\frac{2\pi}{\be}\ze_{2\pi}(s;x|L_D)$. 
So the integral on the right represents 
$\ze_\be(s;x|L_D)-\frac{2\pi}{\be}\ze_{2\pi}(s;x|L_D)$, in agreement with
the second point of view which we have discussed in the previous section.

Now, in order to perform the integral,
in Eq.~(\ref{zzz}) we shift the contour on the right 
and using Eq.~(\ref{seel}) we obtain
\beq
\ze_\be(s;x|L_D)&=&
\frac{\ze(s-1;\vec x|L_N)}{2\be(s-1)}
\nn\\&&
+\frac1{\be\Ga(s)}\sum_{r=0}^{P}
A_r(\vec x|L_N)I_\be(s+r-\fr N2)\rho^{2s+2r-D}
+O(\rho^{2s+2P-D})
\:,\label{zeLoc}\eeq
$P$ being an arbitrary large integer. By means of this equation and 
of inverse Mellin transform, one comes back to the local heat-kernel asymptotic
\beq
K_\be(t;x|L_D)
\sim\frac1{4\pi t}\sum_{r=0}^{\ii} 
A_r(\vec x|L_N) t^{n-\frac N2}\sim\frac1{4\pi t}K(t;\vec x|L_N)
\:.\label{gui}
\eeq

By integrating Eq.~(\ref{zeLoc}) on the manifold with the smeared 
function $\phi(\rho)=\theta(\La-\rho)\theta(\rho-\ep)$ 
one has
\beq
\ze_\be(s|L_D)(\phi)&=&\ze_{2\pi}(s|L_D)
\nn\\&&\hspace{-1cm}
+\frac1{\Ga(s)}\sum_{r=0}^{P}
A_r(L_N)I_\be(s+r-\fr N2)\hat\phi(2s+2r-N)+f(s;\La,\ep)
\:,\label{zef1}
\eeq
where $\ze_{2\pi}(s|L_D)=\frac{\ze(s-1|L_N)\hat\phi(2)}{2(s-1)}$ and
$f(s;\La,\ep)$ is an analytic function in $s$ 
going to 0 as $\ep\to0$.

Now we may write down the meromorphic structure 
of global zeta-function, which can directly read off by looking at 
Eq.~(\ref{zef1}) and recalling that $I_\be(s)$ has only a simple pole at 
$s=1$. The result reads
\beq
\Ga(s)\ze_\be(s|L_D)(\phi)&\sim&\sum_{r=0}^{\ii}
\frac{A_r(L_N)\aq\hat\phi(2)+\at\frac{\be}{2\pi}-1\ct
\hat\phi(2s+2r-N)\cq}{2(s+r-\frac D2)}
\:.\nn\eeq
Taking the limit for $s\to0$ of the latter equation we easily get
$\ze_\be(0|L_D)=0$ for odd $D$, while for even $D$
\beq
\ze_\be(0|L_D)=\frac{\be}{4\pi}A_{D/2}(L_N)\hat\phi(2)
\nn\eeq
and this is just the integral of the $\frac D2$ coefficient in the  
asymptotic expansion (\ref{gui}).
It has to be noted that $\ze_\be(0|L_D)$ is linear in $\be$ 
(or vanishing) and so the $\mu$ dependence in  Eq.~(\ref{FE}),
which reflects the zeta-function ultraviolet renormalisation, does not 
contribute to the entropy and this is again in agreement with the conformal 
transformation method.

\s{Statistical mechanics in Rindler-like space-times}

Now, making use of the expression for the zeta-function we have derived in 
the previous section, Eq.~(\ref{zef1}), 
we  can study the statistical mechanics
for scalar fields in a Rindler-like space-time by means of 
Eq.~(\ref{FE}).
Taking the derivative of Eq.~(\ref{zef1}) and the limit $s\to0$
we obtain
\beq
F_\be&=&-\frac1{2\be}\ze'_{2\pi}(0|L_D)
-\frac1{2\be}\sum_{r=0}^{\ii} 
A_r(L_N)I_\be(r-\fr N2)\hat\phi(2r-N)
\nn\\&&\hs\hs
-\frac{\ze_\be(0|L_D)}{2\be}\:\:\ln\mu^2
-\frac{f(0;\La,\ep)}{2\be}
\:.\nn\eeq
As usual, it is convenient to distinguish between odd and even dimensional 
cases respectively,  i.e.
\beq
F_\be&=&-\frac1{2\be}\sum_{r=0}^{\frac{N-1}2} 
\frac{A_r(L_N)I_\be(r-\frac N2)}{(N-2r)\ep^{N-2r}}
-\frac1{2\be}\ze'_{2\pi}(0|L_D)
+O(\La^2)\:,\label{FEodd}\\
F_\be&=&-\frac1{2\be}\sum_{r=0}^{\frac N2-1} 
\frac{A_r(L_N)I_\be(r-\frac N2)}{(N-2r)\ep^{N-2r}}
-\frac{A_{N/2}(L_N)I_\be(0)}{2\be}\:\ln\frac{\La^2}{\ep^2}
\nn\\&&\hs\hs\hs\hs
-\frac1{2\be}\ze'_{2\pi}(0|L_D)
+O(\La^2)\:.\label{FEeven}\eeq
The last term $O(\La^2)$ contains also the 
dependence on the scale parameter $\mu$. In any case such a 
term is independent on $\be$ and does not give contributions to the 
entropy.

Some comments are in order. We observe that
the free energy in the vicinity of the horizon 
has a number of  divergences depending on the dimension $D$
and in the even dimensional case 
also a logarithmic divergence appears, 
proportional to $A_{N/2}$, in agreement with results 
obtained by other methods \cite{byts96-458-267}.
There is a finite part linear in the temperature and 
proportional to $\ze'_{2\pi}$. Besides,
the leading term has the expected $\be^{-D}$ behaviour. 
This is a non trivial result and it is a consequence of our local 
approach, which requires the analytical continuation
investigated in the appendix. 
It fact, from Eqs.~(\ref{IN2e}) and (\ref{IN2o}) we have
\beq
F_\al &\sim&
-\frac{(-1)^{\frac{D-1}2}\ze'_R(1-D)}
{\sqrt\pi(4\pi)^{\frac D2}\Ga(\frac{D+1}2)}
\:\frac{V_N}{N\ep^N}\:\:\:\al^D\:,
\hs D=3,5,7,...\nn \\
F_\al&\sim&
-\frac{\Ga(\frac{1-D}2)\ze_R(1-D)}{\sqrt\pi(4\pi)^{\frac D2}}
\:\frac{V_N}{N\ep^N}\:\:\:\al^D\:,
\hs D=4,6,...
\nn\eeq
Here $\ze_R$ is the usual Riemann zeta-function and 
$\al=2\pi/\be$. Note that here the equilibrium temperature corresponds to 
$\al=1$.

Using  Eq.~(\ref{entropy}) or the equivalent relation
\beq
S_{\al}=-2\pi\frac{\partial F_\al}{\partial\al}
\eeq
and Eqs.~(\ref{FEodd}-\ref{FEeven}), one can compute the entropy in any 
$D$-dimensional Rindler-like space-time.
For the sake of simplicity, here we shall deal with the 4-dimensional 
case only. One easily gets
\beq
F_\al=-\frac{A(\al^2-1)(\al^2+11)}{180(4\pi)^2\ep^2}
+\frac{A_1(L_2)(\al^2-1)}{48\pi}\ln\frac{\La^2}{\ep^2}
-\frac{\al}{4\pi}\ze'_{2\pi}(0|L_4)+O(\La^2)
\:,\label{FE4}\eeq
$A=V_2$ being the transverse area and 
finally, at the equilibrium temperature $\al=1$
\beq
S_{\al=1}=\frac{A}{60\pi\ep^2}
+\frac{A_1(L_2)}{12}\ln\frac{\La^2}{\ep^2}
+\frac12\ze'_{2\pi}(0|L_4)+O(\La^2)
\:.\nn\eeq

The results which we have obtained in this section are valid for a 
scalar field in a general Rindler-like space-time.
For a massless scalar field in a Rindler space-time,
$\ze_{2\pi}(s|L_D)$, as well as
all coefficients $A_r(L_N)$, but $A_0$ vanish.
So, previous formulae reduce to 
\beq
F_\be=-\frac{I_\be(0)}{2\be}\ln\frac{\La}{\ep}=-\frac{1}{24 \pi} 
\ln\frac{\La}{\ep}\at \al^2-1 \ct
\:,\hs D=2
\:,\nn\eeq
\beq
F_\be=-\frac{I^\be(-N/2)}{2\be(4\pi)^{N/2}}\frac{V_N}{N\ep^N}
\:,\hs D\geq3
\:,\nn\eeq
which  can be directly derived from Eq.~(\ref{mnmr7}).
Finally, for $D=2$ and $D=4$ the entropies read respectively
\beq
S_{\al=1}=\frac{1}{6} \ln\frac{\La}{\ep}
\:,\hs D=2\,\,\,, 
  S_{\al=1}=\frac{A}{60\pi\ep^2} \hs D=4
\:.\label{bu}\eeq
The first reproduce the well known 2-dimensional result, while the 
latter is compatible with the same quantity calculated 
by other methods, the only difference being the numerical factor in 
the denominator.

In this paper we are mainly interested in entropy. However, in the 
following we shall briefly discuss the renormalisation of the internal 
energy. We remind that we have made use of an analytical regularisation. We have 
at disposal a renormalisation prescription.  In this 
approach it is quite natural to require the internal energy 
to be finite (vanishing) 
when the conical singularity is absent, namely when $\be=2\pi$. This can be 
accomplished by making use of the same statistical mechanical 
identities among the renormalised quantities and assuming (for 
example $D=2$)
\beq
U_\al^R=U_\be-U_{\al=1}= \frac{1}{24 \pi}\ln\frac{\La}{\ep}
\aq \al^2-1\cq 
\:.\nn\eeq
This prescription automatically gives 
\beq
S_\al^R=S_\al=\frac{1}{6}\ln\frac{\La}{\ep} 
\:.\nn\eeq
Note that $U_{2\pi}$ depends on the horizon cutoff. 
The corresponding free energy reads
\beq
F_\al^R=\frac{1}{24 \pi}\ln\frac{\La}{\ep}
\aq \al^2-1\cq -\frac{1}{12 \pi}\ln\frac{\La}{\ep}\al^2
\:.\eeq
which is not vanishing for $\be=2\pi$. The same analysis can be extended to 
the higher dimensional cases. 

We conclude this Section with few remarks. 
With regard to other off-shell computations of 
the entropy and free energy for a black hole, we recall that 
the horizons divergences 
can be obtained for example within the path integral approach, 
making use of the 
high-temperature approximation \cite{dowk78-11-895,deal95-52-3529}. This 
gives the correct leading term in $\alpha^4$, proportional to the 
optical volume and this result is in agreement with our expression
of the free energy in the 4-dimensional case. However, in this case 
there is also a disagreement with the computations based on the other 
off-shell methods, occuring in the  $\alpha^2$ terms, and this 
leads to the  anomalous numerical coefficient in  Eq.~(\ref{bu}) 
for the expression of the entropy. 

\s{$D$-dimensional black hole near the horizon and in the large mass limit}

Here we consider  the case in which $\M^D=C_\be\X S^N$.
To justify this choice from a physical view point,
first of all we show that, near the horizon and in the large black 
hole mass, a (Euclidean)
$D$-dimensional black hole with vanishing cosmological constant,
may be approximated by a manifold of this kind and so, 
the statistical mechanics can be investigated by using the formulae of previous 
Sections. 

We recall that the static metric describing a $D$-dimensional Schwarzschild 
black hole (we assume $D>3$ and vanishing cosmological constant) 
read \cite{call88-311-673} 
\beq
ds^2=-\aq1-\at\frac{r_H}r\ct^{D-3}\cq\,dx_0^2+
\aq1-\at\frac{r_H}r\ct^{D-3}\cq^{-1}\,dr^2
+r^2\,d\Omega_{D-2}
\:,\nn\eeq
where we are using polar coordinates, $r$ being the radial one and 
$d\Omega_{D-2}$ the $D-2$-dimensional spherical unit metric.
The horizon radius is given by
\beq
r_H=\aq\frac{2\pi^{\frac{D-3}2}MG_D}
{(D-2)\Ga(\frac{D-1}2)}\cq^{\fr1{D-3}}
\:,\nn\eeq
$M$ being the mass of the black hole and $G_D$ 
the generalised Newton constant. Tha associated Hawking temperature 
reads  $\be_H=4\pi r_H/(D-3)$. From now on, we put $r_H=1$. 

To study the black hole near the horizon
it is convenient to redefine the radial Schwarzschild coordinates 
$x_0=x_0'/b$ and $r=r(\rho)$ by means of the implicit relation
\beq
\at\frac{b\rho^2}2\ct^{\fr1{D-3}}
=e^{r-1}\exp\int\frac{dr}{r^{D-3}-1}\:,\hs
\rho^2&\sim&\frac{1-r^{3-D}}{b^2}
\:,\nn\eeq
where $b=(D-3)/2$.
In the new set of coordinates we have
\beq
ds^2=-\frac{1-r^{3-D}(\rho)}{b^2}\,dx_0'^2+
\frac{1-r^{3-D}(\rho)}{b^2\rho^2}\,d\rho^2
+r^2(\rho)\,d\Omega_{D-2}
\eeq
and finally, near the horizon
\beq
d s^2 =-\rho^2dx_0'^2+d\rho^2+\,d\Omega_N 
\:,\nn\eeq
which is the metric of a Rindler-like space $C_\be\times S^N$. 
As a consequence we can use all results developed in previous sections. 
In particular, the zeta-function can be computed making use of 
Eq.~(\ref{zzz}), since  
the zeta-function for the Laplace operator on $ S^N$ is well known 
(see for example \cite{camp90-196-1,byts96-266-1}). 
More simply, we can directly
derive the free energy by using Eqs.~(\ref{FEodd}-\ref{FEeven}).

For its physical interest now we shall investigate 
in more detail the case $D=4$,
a toy model for the 4-dimensional eternal black hole. 
For this case we have
\beq
\Ga(z)\ze(z|L_2)=2\sum_{k=0}^\ii 
\frac{(-a^2)^k}{k!} \Ga(z+k)\ze_H(2z+2k-1;\fr{1}{2})
\:,\nn\eeq
where $L_2=-\lap_{S^2}+\fr{1}{4}+a^2$ 
and $\ze_H(s;q)$ the Hurwitz zeta-function. 
All the spectral coefficients may be evaluated from the above 
expression computing the residues at the simple poles $z=1-r$, 
($r=0,1,2,...$). As a result $A_0=1$ and
\beq
A_r(L_2)=(-1)^r\aq\frac{a^{2r}}{r!}
-2\sum_{j=0}^{r-1}
\frac{a^{2j}\ze_H(2j-2r+1;\fr12)}{j!(r-j-1)!}
\cq\:,\hs r\geq1
\:.\nn\eeq

The free energy is given by Eq.~(\ref{FE4}) and reads
\beq
F_\al&=&-\frac{A(\al^2-1)(\al^2+11)}{180(4\pi)^2\ep^2}
+\at\frac{1}{12}-a^2\ct
\frac{(\al^2-1)}{48\pi}\ln\frac{\La^2}{\ep^2}
\nn\\&&\hs\hs\hs
-\frac{\al}{4\pi}\ze'_{2\pi}(0|L_4)+O(\La^2)
\:,\nn\eeq
where $A=4\pi r_H^2$ is the horizon area. 
Also in this case, we may require that the internal energy has to be finite 
at the Hawking temperature. This can realised adding the infinite 
constant $-U_{2\pi}$. However, this prescription does not modify the 
entropy.
For a massless scalar field one has $a^2=-1/4$ and so
the (renormalised) entropy at the equilibrium temperature $\al=1$ 
(which means $\be_H=2\pi r_H/b=8\pi MG$) is
\beq
S_{\al=1}=\frac{A}{60\pi\ep^2}
-\frac1{36}\ln\frac{\La^2}{\ep^2}
+\frac1{2}\ze'_{\R^2\X S^2}(0|L_4)+O(\La^2)
\:.\nn\eeq

It should be noted the appearance of the 
logarithmic horizon divergences \cite{solo95-51-609,cogn95-12-1927}, which is absent in the massless 
Rindler case. Higher dimensional cases can be analysed on the same 
lines, without any difficulties.

\s{Conclusions}

In this paper the  entropy for a massless 
scalar field in a D-dimensional Rindler-like space-time has been 
investigated by means of the off-shell conical Euclidean method based 
on a local zeta-function regularisation. The 
degrees of freedom of the black hole has been assumed to be  equivalent to 
those of a massless scalar quantum field on 
$C_\be\times\M^N$, $C_\be$ being the two dimensional cone.  
The period  $\be$ of the imaginary compactified time 
has been interpreted as the inverse of the temperature measured at infinity. 
One of the advantage of this approach is the determination of the 
unique equilibrium temperature, the Unruh-Hawking temperature, by 
the requirement of the absence of conical singularities ($\be=2\pi$). 
Within this approach, the Bekenstein-Hawking entropy can also be 
obtained, but again without a statistical interpretation. 
With regard to this issue, the formal partition 
function related to the determinant of a Laplace-like operator 
on  $C_\be\times\M^N$, evaluated off-shell ($\be\neq2\pi$), in order to 
permit the computation of the entropy by means of statistical 
formulae, has been regularised according to the zeta-function 
method. This has posed the mathematical  problem of defining 
properly the related zeta-function. 

A suitable analytical procedure 
first suggested by Cheeger has been used in order to implement the 
usual relationship between local zeta-function and heat-kernel, as 
well as the corresponding traces, for which a smearing 
function has been introduced in order to define them.  
The so called horizon divergences of this  
entropy evaluated at the equilibrium temperature, 
which are also present if one is dealing 
with other off-shell techniques, are recovered 
with a natural choice of the smearing function. We have obtained 
agreement with other methods (the conformal transformation method, 
the brick-wall model and the canonical approach), 
the only difference being the 
numerical coefficients of the divergences, even 
thought the structural form of the divergent terms is the same.

Another by-product of our conical Euclidean approach has been 
the dimensionally correct leading behaviour of the free energy in $\be$, 
a result that the global heat-kernel method completely misses. In the usual 
conical approach, the one we have called global approach, 
one needs a mass as infrared cut-off and has to use 
proper-time regularisation, namely an ultraviolet regularisation 
different from zeta-function. 
The leading divergence survives in the 
limit $m\to0$, but one gets again a  $\be$ behaviour 
independent on the dimensions of the manifold \cite{kaba95-453-281}. 
Furthermore, within this approach these 
"ultraviolet" divergences are interpreted as horizon divergences. 
As a result, 
even though the interpretation of the divergences is 
different, the conclusions are similar. With regard to this issue, 
one should try to investigate the limit $m \to 0$ and the infinite 
volume limit by starting, ab initio, with a truncate cone and 
imposing suitable boundary conditions. Then 
one should study the massless limit and the infinite cone limit in 
order to better understand the existence of an infrared phenomena. 
Our local approach has implicitly assumed the infinite cone limit.
We stress again that the Cheeger method permits to study the massless 
case directly in the infinite cone case.   

As far as the horizon divergences of the off-shell quantities are 
concerned, we have little to add to the considerations 
recently appeared  in the literature and a detailed discussion 
can be found in Ref.~\cite{frol95u-184}. 
There it has been shown, working with two 
dimensional models, that all the observables related to a black 
hole at Hawking temperature can be evaluated in terms of on-shell 
finite quantities and a subtraction procedure between the on-shell and 
off-shell quantities has been proposed, the divergences of the former 
being removed by the related quantities in the Rindler space-time.     
Another proposal to deal with such divergences, consisting in the 
implementation of the 't~Hooft approach by means of Pauli-Villars 
regularisation, has been recently introduced in 
Ref.~\cite{deme95-52-2245} 
and it has been used in a 2-dimensional model 
in Ref.~\cite{solo96u-154}, where a comparison 
between the Frolov-Fursaev-Zelnikov  scheme 
\cite{frol95u-184} and the latter can be found.
The absence of the on-shell entropy divergences  has been also claimed 
in Ref.~\cite{belg95u-521}. 
 
We conclude with some remarks. The Rindler case could be the key example in order to 
better understand the horizon divergences. Here it is well 
established that the internal energy must be finite (actually vanishing) 
if and only if $\be=\be_H$. 
As a consequence, the related entropy is divergent at 
the same equilibrium temperature. It has been shown that such 
statistical-mechanical entropy
coincides with the entropy of entanglement obtained from the density 
matrix describing the vacuum state of the field 
(scalar or spinor) as observed from one 
side of a boundary in Minkowski space-time  
\cite{bomb86-34-373,sred93-71-666,call94-333-55,kaba94-329-46,kaba95-453-281}.  
However the entropy of entanglement, although formally divergent, 
might be operationally finite \cite{beke95b}.  With regard 
to the black hole case,  the fluctuations of the horizon 
\cite{thoo85-256-727} as a well as the 
quantum evaporation \cite{russ95u-9} 
might provide again a mechanism for the absence of the entropy divergences.
Finally, we have to mention that recently several 
attempts to clarify the microscopic 
origin of black hole entropy have been appeared within the string theory, 
which seems, at the moment, a  promising theory capable 
to offer a solution of this important
issue (see for example \cite{stro96u-129} and references therein).  

\ack{We would like to thank P. Menotti and P. Peirano for useful 
discussions.}

\appendix

\s{Properties of the $G_\be(s)$ function}
\label{S:Gal}

Here we make the analytic continuation of the function
\beq
G_\be(s)=\sum_{l=1}^{\ii}\frac{\Ga(\nu_l-s+1)}{\Ga(\nu_l+s)}
\:,\hs\nu_l=\al l
\:,\hs \al(\be)=\frac{2\pi}{\be}
\:,\nn\eeq 
which is convergent for $\Re s>1$. In fact, 
for $\nu\to\ii$ one has the asymptotic expansion
\beq
\frac{\Ga(\nu-s+1)}{\Ga(\nu+s)}\sim
\nu^{1-2s}\sum_{j=0}^{\ii}c_j(s)\nu^{-2j}
\:,\nn\eeq
where $c_j(s)$ are easily computable using the known expansion of 
$\Ga(z)$, that is
\beq
\Ga(z)\sim\sqrt{\frac{2\pi}z}e^{-z+z\ln z+B(z)}
\:,\hs B(z)=\sum_{j=0}^{\ii}\frac{B_{2j}z^{1-2j}}{2j(2j-1)}
\:,\nn\eeq
$B_j$ being the Bernoulli numbers.
It it easy to see that since the function 
$\frac{\Ga(\nu-s+1)}{\Ga(\nu+s)}$ for any $s=-n/2$ ($n=-1,0,1,2,...$) 
is effectively a polynomial of order $\nu^{n+1}$, 
$c_j(-n/2)$ has to be vanish for all $j>(n+1)/2$. 

The first coefficients read $c_0(s)=1$,
\beq
c_1(s)=\frac{s(s-1/2)(s-1)}3;\hs 
c_2(s)=\frac{s(s^2-1/4)(s^2-1)(s-6/5)}{18}
\:.\label{GjCoeff}\eeq

It has to be noted that $G_\be(s)$ is certainly analytic in the strip 
$1<\Re s<1+\nu_1$ and, as we shall see later, it
has a simple pole at $s=1$ with residue equal to $1/2\al$. 
In order to make the analytic continuation of $G_\be(s)$ we define
\beq
f_n(\nu,s)=\frac{\Ga(\nu-s+1)}{\Ga(\nu+s)}
-\sum_{j=0}^{\aq\frac{n}2\cq+1}c_j(s)\nu^{1-2s-2j}
\sim c_{\aq\frac{n}2\cq+2}(s)
\nu^{-\at2s+2\aq\frac{n}2\cq+3\ct}
\:,\nn\eeq
where $\aq\frac{n}2\cq$ represents the integer part of $\frac{n}2$. 
For $\Re s>1$ we have
\beq
G_\be(s)=\sum_{j=0}^{\aq\frac{n}2\cq+1}\al^{1-2s-2j}
c_j(s)\:\ze_R(2s+2j-1)+\sum_{k=1}^{\ii}f_n(\nu_k,s)
\:,\label{Gs}\eeq
Now, the right hand side of the latter equation has meaning for
$\Re s>-1-\aq\frac{n}2\cq$ and so we have obtained the analytic 
continuation we were looking for.
The function $f_n(\nu,s)$ is in general unknown, but it is vanishing
for all $s=1/2,0,-1/2,-1,...,-n/2$, since in this case
the function $\frac{\Ga(\nu+1+n/2)}{\Ga(\nu-n/2)}$ is a polynomial.
Then for $n=-1,0,1,2,3,...$ we obtain
\beq
G_\be(-n/2)&=&\sum_{j=0}^{\aq\frac{n}2\cq}
\al^{n+1-2j}c_j(-n/2)\:\ze_R(2j-n-1)
\nn\\&&\hs\hs
+\ap\al^{n-1-2\aq\frac{n}2\cq}c_{\aq\frac{n}2\cq+1}(s)
\:\ze_R(2s+2\aq\fr{n}2\cq+1)\right|_{s=-n/2}
\:.\label{Gn2}\eeq

Using Eqs.~(\ref{GjCoeff}), (\ref{Gs}) and (\ref{Gn2}) we have
\beq
\ap\Res G_\be(s)\right|_{s=1}=\frac\be{4\pi}
\:,\hs
G_\be(0)=\frac1{12}\at\frac{1}{\al}-\al\ct\;,
\eeq
\beq
G_\be(-1)=\frac1{120}\at\al^3+10\al-\frac{11}{\al}\ct
\:.\eeq
Recalling that $\ze_R(0)=-1/2$ and $\ze_R(-2j)=0$ for any $j\in\N$, 
from Eq.~(\ref{Gn2}) we also get
\beq
G_\be(-n/2)=-\fr12c_{\frac{n+1}2}(-n/2)
=-\frac{\Ga(1+n/2)}{2\Ga(-n/2)}
\:,\hs n=-1,1,3,5,...
\:.\label{Gn22}\eeq
The latter expression has been derived from the identity
\beq
\frac{\Ga(\nu+n/2+1)}{\Ga(\nu-n/2)}
=\aq\nu^2-\at\frac12\ct^2\cq
\aq\nu^2-\at\frac32\ct^2\cq \cdots
\aq\nu^2-\at\frac{n}2\ct^2\cq
\:,\nn\eeq
valid for any odd $n=-1,1,3,5,...$.
The identity
\beq
G_{2\pi}(s)=-\frac{\Ga(1-s)}{2\Ga(s)}
\:,\nn\eeq
also holds. From the latter equation and Eq.~(\ref{Gn22}) we have
\beq
G_\be(-n/2)-G_{2\pi}(-n/2)=0\:,\hs\hs n=-1,1,3,5,...
\:.\nn\eeq

In the paper we frequently meet the function
\beq
I_\be(s)=\frac{\Ga(s-1/2)}{\sqrt\pi}
\aq G_\be(s)-G_{2\pi}(s)\cq
\:,\label{Ibe}\eeq
which has a simple pole at $s=1$. We have
\beq
\ap\Res I_\be(s)\right|_{s=1}=
\frac12\at\frac\be{2\pi}-1\ct\:,\hs
I_\be(0)=-2G_\be(0)=\frac16\at\frac\be{2\pi}-\frac{2\pi}\be\ct
\nn\eeq
and by definition $I_{2\pi}(s)=0$. 
We also need the behaviour of $I_\be$ with respect to 
$\be$ at $s=-N/2$. 
From Eqs.~(\ref{Gn2}) and (\ref{Ibe}) for
even $N$ we immediately have
\beq
I_\be(-N/2)&=&\frac{\Ga(-\frac{N+1}2)}{\sqrt\pi}
\aq G_\be(-N/2)-G_{2\pi}(-N/2) \cq\nn\\
&\sim&\frac{\Ga(-\frac{N+1}2)\ze_R(-N-1)}{\sqrt\pi}
\:\:\:\al^{N+1}\:,
\hs N=0,2,4,...
\:,\label{IN2e}\eeq
while for odd $N$, using Eq.~(\ref{Gs}) we obtain
\beq
I_\be(-N/2)&=&\frac{(-1)^{\frac{N+1}2}}
{2\sqrt\pi\Ga(\frac{N+3}2)}
\aq G'_\be(-N/2)-G'_{2\pi}(-N/2)\cq\nn\\
&\sim&\frac{(-1)^{\frac{N+1}2}\ze'_R(-N-1)}
{2\sqrt\pi\Ga(\frac{N+3}2)}
\:\:\:\al^{N+1}\:,
\hs N=1,3,5,...
\label{IN2o}\eeq
In the evaluation of the latter expansion, we have considered only the 
first term on the righ-hand side of Eq.~(\ref{Gs}), since the derivatives 
at $s=-N/2$ of the functions $f_N(\nu_k,s)$ 
give contributions of the order $\al^{-2}$.

\end{document}